\documentclass[12pt, preprint]{aastex}

\shorttitle{Black Hole masses and the Fundamental Plane  of BL Lacs}
\shortauthors{Falomo et al.}

\def\ref{\par \noindent \hang}
\def\emu{{< \hspace{-4pt} \mu_e \hspace{-4pt}>}}
\def\mincir{\ \raise -2.truept\hbox{\rlap{\hbox{$\sim$}}\raise5.truept
\hbox{$<$}\ }}

\def\CMcal{}

\begin{document}

\title{Black hole masses and the Fundamental Plane of BL Lac objects}

\author{R. Falomo}
\affil{INAF -- Osservatorio Astronomico di Padova,  Vicolo dell'Osservatorio 5,  
35122 Padova,  Italy}
\email{falomo@pd.astro.it}

\author{J.K. Kotilainen}
\affil{Tuorla Observatory,  University of Turku,  V\"ais\"al\"antie 20,  
FIN--21500 Piikki\"o,  Finland}
\email{jarkot@astro.utu.fi}

\author{N. Carangelo}
\affil{Istituto di Astrofisica Spaziale e Fisica Cosmica (IASF),  
Sez. di Milano - CNR,  Via Bassini,  15,  I-20133 Milano,  Italy}
\email{nicol@mi.iasf.cnr.it}

\and

\author{A. Treves}
\affil{Universit\`a dell'Insubria,  via Valleggio 11,  22100 Como,  Italy}
\email{treves@mib.infn.it}

\begin{abstract} 
We report on measurements of the stellar velocity dispersion ($\sigma$) from 
the optical spectra of the host galaxies of four BL Lac objects. Together with 
our earlier results on seven BL Lac objects (Falomo et al. 2002),  
and with the previously derived photometrical and structural properties,  
these data are used to construct the Fundamental Plane (FP) of the 
BL Lac hosts. We find that the BL Lacs follow the same FP as low redshift 
radio galaxies and inactive luminous ellipticals,  
in agreement with similar results presented by Barth et al (2003). This indicates that the 
photometrical,  structural and kinematical properties of the host galaxies of 
BL Lacs are indistinguishable from those of inactive massive ellipticals.  
Using the correlation between black hole mass ($M_{BH}$) and $\sigma$ in 
nearby elliptical galaxies,  we derive the masses of the central black hole in 
BL Lacs. These masses,  in the range of 6 $\times$ 10$^7$ to 
9 $\times$ 10$^8$ M$_\odot$,  are consistent with the values derived from the 
bulge luminosity and appear to be linearly correlated with the mass of the 
galaxies ($M_{BH}$ $\approx$ 0.001$\times$ $M_{bulge}$).
\end{abstract}


\section{Introduction}

Imaging studies of the host galaxies of low redshift BL Lac objects (e.g. 
Falomo 1996; Wurtz,  Stocke \& Yee 1996; 
Falomo \& Kotilainen 1999; Urry et al. 2000) have consistently shown that this kind of 
nuclear activity is exclusively hosted by luminous elliptical galaxies. 
Detailed studies of the nearest (z $<$ 0.2) objects have indicated that the 
structural properties of the hosts of these AGN are indistinguishable from 
those of normal unperturbed massive ellipticals (Falomo et al. 2000). 
Similar studies conducted on a large sample of nearby radio galaxies 
(Govoni et al. 2000) have shown them to have also photometrical and structural 
global properties close to those of inactive spheroids. These observations 
lend support to the unified model for radio-loud AGN 
(e.g. Urry \& Padovani 1995) that interpret BL Lac objects as a subset of 
radio galaxies with the jet pointing close to our line-of-sight.

On the other hand,  it is also well known that the global properties of 
early--type galaxies,  such as the effective radius $r_e$,  the average 
surface brightness $\emu$ and the central velocity dispersion $\sigma_c$ are 
linked fairly well through the so-called Fundamental Plane (hereafter FP; 
Djorgovski \& Davis 1987,  Dressler et al. 1987). This is in fact what is 
expected if galaxies are virialized systems,  have constant M/L ratio and are 
homologous structures.

Indeed,  real galaxies deviate from the FP defined by the virial
equilibrium under the above hypotheses (e.g. Burstein et al. 1997; J\o
rgensen et al. 1996) and define a plane that is tilted with respect to
the virial FP.  The differences from the 'canonical' FP are
interpreted as due to changes in the M/L ratio (due to different
stellar content or differences in the dark matter distribution; Pahre, 
Djorgovski \& de Carvalho 1995; Mobasher et al. 1999),  or to the
breakdown of the homology assumption (e.g. kinematic anisotropy; Ciotti,  Lanzoni \& Renzini 1996;
Busarello et al. 1997).

In the context of the link between galactic properties and the nuclear 
activity,  it is important to understand if active galaxies define the same FP 
as normal (inactive) ellipticals. A recent study of a large sample of 
powerful radio galaxies (Bettoni  et al 2001) has provided clear evidence that this is 
the case. The host galaxies of these radio sources define in fact an FP that 
is fully consistent with that of inactive ellipticals with the only difference 
of sampling the most luminous (and massive) objects of the luminosity function 
of early type galaxies.

In this paper we present new measurements of the stellar velocity dispersion 
($\sigma$) in the host galaxies of four BL Lac objects that together with our 
earlier results on seven BL Lac objects (Falomo, Kotilainen \& Treves 2002, hereafter FKT02),  
and with the previous determinations of the global photometrical and 
structural host properties,  are used to construct the FP of BL Lacs and 
to compare it with that of radio galaxies and normal ellipticals.

In addition,  we make use of the correlation relating the central black
hole (BH) mass ($M_{BH}$) with $\sigma$ of the spheroidal component in
nearby inactive galaxies (e.g. Gebhardt et al. 2000; Ferrarese \&
Merritt 2000) to infer the $M_{BH}$ of our sample of BL Lacs thus
expanding our previous study (FKT02). Similar
measurements of the velocity dispersion of BL Lacs (with many objects
in common with our sample) have been reported recently by Barth,  Ho \&
Sargent (2003; hereafter BHS03).  Preliminary results of this work are given in
Treves et al. (2003) and Falomo et al. (2003).
Hubble constant H$_0$ = 50 km s$^{-1}$
Mpc$^{-1}$ and $\Omega$ = 0  are used throughout this paper.

\section{Observations and data analysis}

The spectra were obtained in June 2001 using the ESO 3.6m telescope
equipped with EFOSC2. Two grisms were used covering the spectral
ranges 4400 -- 6300 \AA ~(setup A)  and 6300 -- 8000 \AA ~ (setup B) at 0.54 \AA
~pixel$^{-1}$ and 1.3 \AA ~pixel$^{-1}$ dispersion,  respectively.  The
absorption lines from the host galaxies which are considered in our analysis 
comprise of 
H$\beta$ (4861 \AA),  Mg I (5175 \AA),  Ca E-band (5269 \AA),  Na I (5892
\AA) and the TiO + CaI (6178 \AA) and TiO + FeI (6266 \AA) blends.

The grisms and a 1\arcsec\ slit yield a spectral resolution 
of $\sim$60 -- 80 km s$^{-1}$,  which is adequate for the expected range of 
$\sigma$ in luminous ellipticals (e.g. Djorgovski \& Davis 1987; 
Bender,  Burstein \& Faber 1992) such as the hosts of BL Lacs. 
In addition,  spectra of bright KI-KIII  stars that exhibit a low rotational 
velocity (V$\times$$\sin{(i)}<$20 km s$^{-1}$) were taken to be used as 
templates of zero velocity dispersion. Furthermore,  spectra of the
nearby elliptical galaxies NGC 2986,  NGC 5831 and NGC 7562 were secured to 
provide a test of the adopted procedure to derive $\sigma$.

During the observations,  seeing ranged between 0.8\arcsec\ and
1.3\arcsec\ .  The slit was positioned on the nucleus and the 1D
spectrum was extracted from a 3\arcsec\ - 5\arcsec\ diameter aperture, 
which was always within the effective radius of the host galaxy in
order to minimize aperture corrections..  The {\sc IRAF}\footnote{{\sc 
IRAF} is distributed by the National Optical Astronomy 
Observatories,  which are operated by the Association of Universities 
for Research in Astronomy,  Inc.,  under cooperative agreement with 
the National Science Foundation.} package was used for data 
reduction,  including bias subtraction,  flat-fielding,  wavelength
calibration and extraction of 1D spectra. For each source,  two spectra
were combined to remove cosmic ray hits and other occasional spurious
signals in the detector.  In Table 1 we give the observed sources,  the
instrumental setup,  exposure times,  and the S/N of  final
spectra that are reported in Fig. 1 (also for sources studied in FKT02).

The stellar velocity dispersion $\sigma$ was determined using the
Fourier Quotient method (e.g. Sargent et al. 1977).  The spectra were
normalized by subtracting the continuum,  converted to a logarithmic
scale and multiplied by a cosine bell function that apodizes 10\% of
the pixels at each end of the spectrum. 
For all the new observed objects but for PKS 0548--32 the optical spectrum shows emission lines 
(mainly H${\beta}$,  H${\alpha}$ and [OIII] 5007 \AA). These features 
have been removed by a linear interpolation of the adjacent continuum. 
Finally,  the Fourier
Transforms of the galaxy spectra were divided by the Fourier
Transforms of template stars and $\sigma$ was computed from a
$\chi^{2}$ fit with a Gaussian broadening function (see e.g. Bertola
et al. 1984; Kuijken \& Merrifield 1993).  The scatter of the $\sigma$
measurements using different template stars was typically $\sim$10
km$s^{-1}$ and can be considered as the minimum uncertainty. Since
early-type galaxies exhibit gradients in velocity dispersion (Davies
et al 1983,  Fisher,  Illingworth \& Franx 1995),  the measured values of
$\sigma$ have been referred to a common aperture adopting the
procedure described in Falomo et al. (2002).  The observed values of
$\sigma$ and their estimated errors are given in Table 1,  and the
results for the full sample (including those from FKT02 
are given in Table 2.  For the three nearby galaxies we find a good
agreement with the previously published values of $\sigma$ (Prugniel
et al 1998).

For three objects we have spectra in both spectral ranges. The
resulting values of $\sigma$ are in all cases in good agreement,  with
average difference 12 km s$^{-1}$,  ensuring sufficient homogeneity of
data taken with different grisms and/or resolution. For PKS 2201+04 the 
present results are very similar to those 
obtained by us at the NOT (FKT02). For the following discussion 
the average of the pair of measurements was thus used.
 
\subsection{Comparison with Barth et al. (2003) results}

BHS03 have recently reported measurements of $\sigma$
for 11 BL Lac objects. Nine of them are in common with our sample,  and
it is therefore enlightening to directly compare the results.  For
some common objects the $\sigma$ values of BHS03 are derived
from the near-infrared CaII triplet lines,  while our measurements
refer to the visual-red spectral region,  including the g-band,  (4304
\AA),  Mg I (5175 \AA),  Ca E-band (5269 \AA),  Na I (5892 \AA) and the
TiO + CaI (6178 \AA),  TiO + FeI (6266 \AA) at rest frame.  Moreover, 
BHS03 use a different method (direct spectral fitting) from 
ours to derive $\sigma$. These differences,  and the use of different
templates,  can explain modest ($\leq$ 20 km/s) differences in
$\sigma$. Larger differences are more difficult to give reasons for.  
BHS03 assert that their measurements are more accurate because of higher
S/N of their data and the possibility that our measurements are biased
by interstellar absorption and metallicity effects. While it is
certainly true that BHS03 acquired higher S/N,  we believe that
the Na I interstellar absorption in the host galaxies has little,  even
negligible,  effect since all the elliptical host galaxies are likely
to be gas-poor. In the cases where we have data in both spectral
ranges (including and excluding the Na I line),  we find no significant
difference in the derived values of $\sigma$. Galactic Na I absorption
occurs at different wavelength than in the (redshifted) BL Lac hosts
and therefore does not contribute to the broadening of the intrinsic
Na I line. Moreover,  the elimination of emission lines by a linear
continuum interpolation has no effect on the Fourier analysis since it
introduces low spatial frequencies that are removed in the Gaussian
fitting of the real part of the power spectrum.  On the other hand, 
metallicity effects on the strength of the Mg I feature are more
challenging to take into account. Note,  however,  that most of the
$\sigma$ data used to derive the M$_{BH}$ -- $\sigma$ relation for
ellipticals were obtained in the visible region (e.g. Davies et
al. 1987),  including the Mg I feature,  and using a method very similar
to ours. We believe that to derive BH masses it is paramount to have
consistent measurements for the targets and the calibrators.

While a good agreement is found for most objects,  
some discrepancies remain. The largest differences of $\sigma$ between 
BHS03 and this work are for Mrk 501 and I Zw 187,  up to 80 km/s and in 
opposite directions. For Mrk 501 (see also FKT02),  our value of 
$\sigma$ is consistent with that expected for the luminosity of the 
host galaxy and gives consistent results for the BH mass (see Fig. 2). For I Zw 187,  
we obtained only a relatively poor S/N spectrum and therefore our 
determination of $\sigma$ remains more uncertain. The influence of these 
differences in $\sigma$ for estimating the BH masses is discussed in the 
next section. 

\section{Results and discussion}

\subsection{Black hole masses of BL Lacs}

Unlike most AGN,  BL Lac objects are characterized by the weakness or absence 
of emission lines. This prevents us from deriving an estimate of the central 
BH mass from the kinematics of regions that are gravitationally bound to 
the BH,  as in Seyfert galaxies and quasars (e.g. McLure \& Dunlop 2001).
A powerful (even if indirect) way to obtain BH masses is thus to use the 
relationship between $M_{BH}$ and $\sigma$ for nearby early-type galaxies. 
Since all sources considered here are at $z \simeq$ 0.1 we assume that possible 
cosmological evolution of the $M_{BH}$ -- $\sigma$ relation is negligible.
We have adopted the updated relationship (derived only from ellipticals) 
by Bettoni et al. (2003) who assume the same cosmology used here and the same 
calibrations for the velocity dispersion and the magnitudes. The relation 
between $\CMcal{M}_{BH}$ and $\sigma$ is: 

\begin{equation}
\log (\CMcal{M}_{BH}/\CMcal{M}_\odot) = 4.55 \log~\sigma -2.27.
\end{equation}

We assume that this relationship is valid also for AGN 
(e.g. Merritt \& Ferrarese 2001) and in particular for BL Lacs which have 
elliptical hosts (e.g. Falomo \& Kotilainen 1999; Falomo et al. 2000; 
Urry et al. 2000). The derived values of $M_{BH}$ are given in Table 2,  
They are in the range of 
$\sim 6\times10^7$ to $\sim 9\times10^8\CMcal {M}_{\odot}$. 
The typical error (median value) on these estimated BH masses,  taking into account the  
uncertainties in $\sigma$ and in the M$_{BH}$ -- $\sigma$ 
relationship given by Bettoni et al (2003)  
is $\sim$ 0.15dex (see Table 2).

The $M_{BH}$ is also correlated with the luminosity of the bulge of the 
host galaxy (e.g. Kormendy \& Gebhardt 2001). 
The absolute magnitudes M$_R$ of the BL Lac hosts are given 
in Table 2. $M_{BH}$ was thus calculated following the updated relationship 
for ellipticals (Bettoni et al. 2003):

\begin{equation}
\log (\CMcal{M}_{BH}/\CMcal{M}_{\odot}) = -0.50 M_{R} - 3.00.
\end{equation}

The resulting values of $M_{BH}$ are given in Table 2. For most sources,  
the difference of $M_{BH}$ derived from the two methods is within the 
estimated uncertainty but for two objects ( PKS 2201+04 and EXO 0706.1+5913) 
the difference is $\sim$ 0.5dex. 

The average values of $M_{BH}$ for the BL Lacs from the two methods are: 
$<\log M_{BH}>_{\sigma}$ = 8.57 $\pm$ 0.12 and 
$<\log M_{BH}>_{bulge}$ = 8.63 $\pm$ 0.08 with an average difference of  
$<\Delta\log M_{BH}>$ -0.07 $\pm$ 0.09. On the other hand,  if we use the 
$\sigma$ values from BHS03 we find:
$<\log M_{BH}>_{\sigma}$ = 8.48 $\pm$ 0.14 and 
$<\log M_{BH}>_{bulge}$ = 8.68 $\pm$ 0.08 with an average difference of  
$<\Delta\log M_{BH}>$ -0.21 $\pm$ 0.10.

In Fig. 2 we compare the BH masses derived from the M$_{BH}$ - M(bulge) 
relation with those obtained from our adopted  M$_{BH}$ - $\sigma$ relation,  using both 
ours and BHS03 values of $\sigma$. There is a generally good 
agreement between the two methods,  although the data from BHS03 tend to 
systematically deviate to lower BH masses. This suggests that some systematic 
effect may be present in the $\sigma$ values obtained by BHS03. 

According to the shape of their spectral energy distribution,  BL Lacs
are distinguished into two types: low frequency peaked (LBL) and high
frequency peaked (HBL; see Giommi \& Padovani 1995). Based on BH
masses derived from the peak luminosity and its frequency,  Wang,  Xue
\& Wang (2001) proposed that LBLs have significantly (by $\sim$ 2dex)
smaller BH masses than HBLs.  In our sample,  there are nine HBLs and
only 3 LBLs (marked as H or L in Table 2). With the caveat of small
number statistics,  we find no significant difference of $M_{BH}$ in
the two types of BL Lacs. This result is confirmed by the analysis of
$M_{BH}$,  derived from host luminosity,  for a larger sample of BL Lacs
(Falomo, Carangelo \& Treves 2003).
 
The measurements of $\sigma$ and the effective radii of the host galaxies can 
be used to estimate the mass of the hosts through the relationship 
(Bender et al 1992)  :

$M_{host}$ = 5$\sigma^{2}r_e$/G.   \hfill (3) 

The two quantities $M_{BH}$ and $M_{host}$ are consistent with a
linear relation and their average ratio is $<M_{BH}$/$M_{host}>$ = 1.0
$\times$ 10$^{-3}$. This is very similar to values found for other
types of active and inactive galaxies (e.g. McLure \& Dunlop 2002;
Bettoni et al. 2003) and indicates that the mass of the spheroid is
fundamentally linked to the mass of the central BH.

\subsection{The fundamental plane of BL Lacs}

The kinematical measurement of the velocity dispersion of the host
galaxies of BL Lacs allows us to construct the FP of these active
galaxies and to compare it with that obtained for other active and
inactive ellipticals. We use the full sample of 12 BL Lacs for which
$\sigma$ has been measured (this work; FKT02; BHS03). 
Since our spectrum of I Zw 187 has rather low S/N,  we
prefer to adopt for it the value of $\sigma$ reported by BHS03.

For all these BL Lacs,  HST/WFPC2 and/or ground-based images are
available to derive the global properties (luminosity and
scale-length) of the host galaxies. In Table 2 we report the data for
the 12 BL Lacs used to construct the FP. Note that while the apparent
magnitude of the host galaxies derived from ground-based and HST
images agree reasonably well,  the effective radius may be different by
more than a factor of two.  As pointed out also by BHS03,  
the effective radius estimated from HST images tends to be
smaller than that obtained from ground-based data.  This is likely due
to the small field of view of WFPC2 ($\sim$35 arcsec) with respect to
the full size ($\sim$30\arcsec\ --- 40\arcsec\ ) of the most nearby
objects. In fact,  this trend disappears when more distant objects are
considered (e.g. Falomo \& Kotilainen 1999). For the purpose of this 
work we have collected all published data reporting both total
apparent magnitude (in the R band) and the effective radius. 
The adopted quantities for each object are derived 
from the median value of all available data. 
In Table 2 we 
report these values together with the r.m.s. scatter for the effective radius for all 
objects with at least two usable measurements.
From the apparent magnitudes and the effective radii R$_e$ of the host galaxies
we have obtained the average surface brightness $<\mu_e>$ within
R$_e$. All magnitude  values have then been corrected for Galactic extinction
(Schlegel,  Finkbeiner \& Davis 1998),  k-correction (Poggianti 1997)
and cosmological (1+z)$^4$ dimming of the surface brightness.

In Fig. 3 we show the FP of the BL Lacs compared with that of low
redshift radio galaxies (Bettoni et al. 2001) and normal ellipticals
(J\"orgensen,  Franx \& Kjaergaard 1996). This is a homogeneous
comparison because all data sets were analyzed identically.  We find a
remarkable agreement between the FP defined by normal (inactive)
galaxies,  radio galaxies and BL Lac host galaxies. 
A similar result was also obtained by BHS03 who presented 
velocity dispersion measurements for most of the objects  discussed here.
The hosts of BL 
Lacs tend to occupy the intermediate region of the FP,  in agreement
with the fact that the hosts of BL Lacs are luminous ellipticals but rarely at the
bright end of the LF of massive ellipticals (Falomo Carangelo \& Treves 2003). The 
comparison of the kinematical,  photometrical and structural properties
of BL Lac hosts with those of radio galaxies and inactive ellipticals
over the FP shows that their global properties are
indistinguishable. This result strengthens the idea that a massive
elliptical galaxy may undergo a phase of nuclear activity with little
(or negligible) effect on its global structure. Moreover,  it gives
support to the hypothesis that the relations connecting the mass of
the central BH with the velocity dispersion and luminosity of nearby
inactive ellipticals,  can be extended to ellipticals with nuclear
activity. The similarity of the BH masses obtained with the two
techniques for BL Lac objects (previous section), shows that this
picture is becoming self-consistent.

For more powerful active galaxies (e.g. radio-loud quasars),  there are no 
kinematical data available,  so it is not possible to compare the global 
properties of their host galaxies in the FP. We note,  however,  that at least 
at low redshift,  the morphology and the photometrical properties of quasar hosts 
agree well with those of luminous inactive ellipticals 
(e.g. Dunlop et al. 2003).

\section{Conclusions}

We have presented kinematical data for a small sample of the host
galaxies of nearby BL Lacs objects that were previously investigated
through images.  The combined photometric and spectroscopic data allow
us to construct the fundamental plane for this class of active
galaxies and to compare it with that of other elliptical galaxies.

The main result of this work,  together with that by BHS03,  show 
that the host galaxies of BL Lacs 
define a FP that is indistinguishable from that of low $z$ radio galaxies and
normal (inactive) ellipticals.  This further strengthens the idea that the nuclear
activity may occur in all massive spheroids without producing
significant changes in the global structural and kinematical properties
of the galaxies.

In addition,  and consistently with the above result,  we have also shown
that the BH mass derived either from the M$_{BH}$ -- $\sigma$ or
$M_{BH}$ - M$_R$ (bulge) relations yield consistent results. The BH
mass of the majority of BL Lacs being contained in the range 10$^8$
to 10$^9$ M$_\odot$ with no difference between HBL and LBL.

\begin{acknowledgments}

This work has received partial support under contracts COFIN 2001/028773,  COFIN 2002/
027145 and 
ASI-IR-35. JKK acknowledges the support of Academy of Finland during part of 
this work.

\end{acknowledgments}

\begin{figure}
\epsscale{1}
\plotone{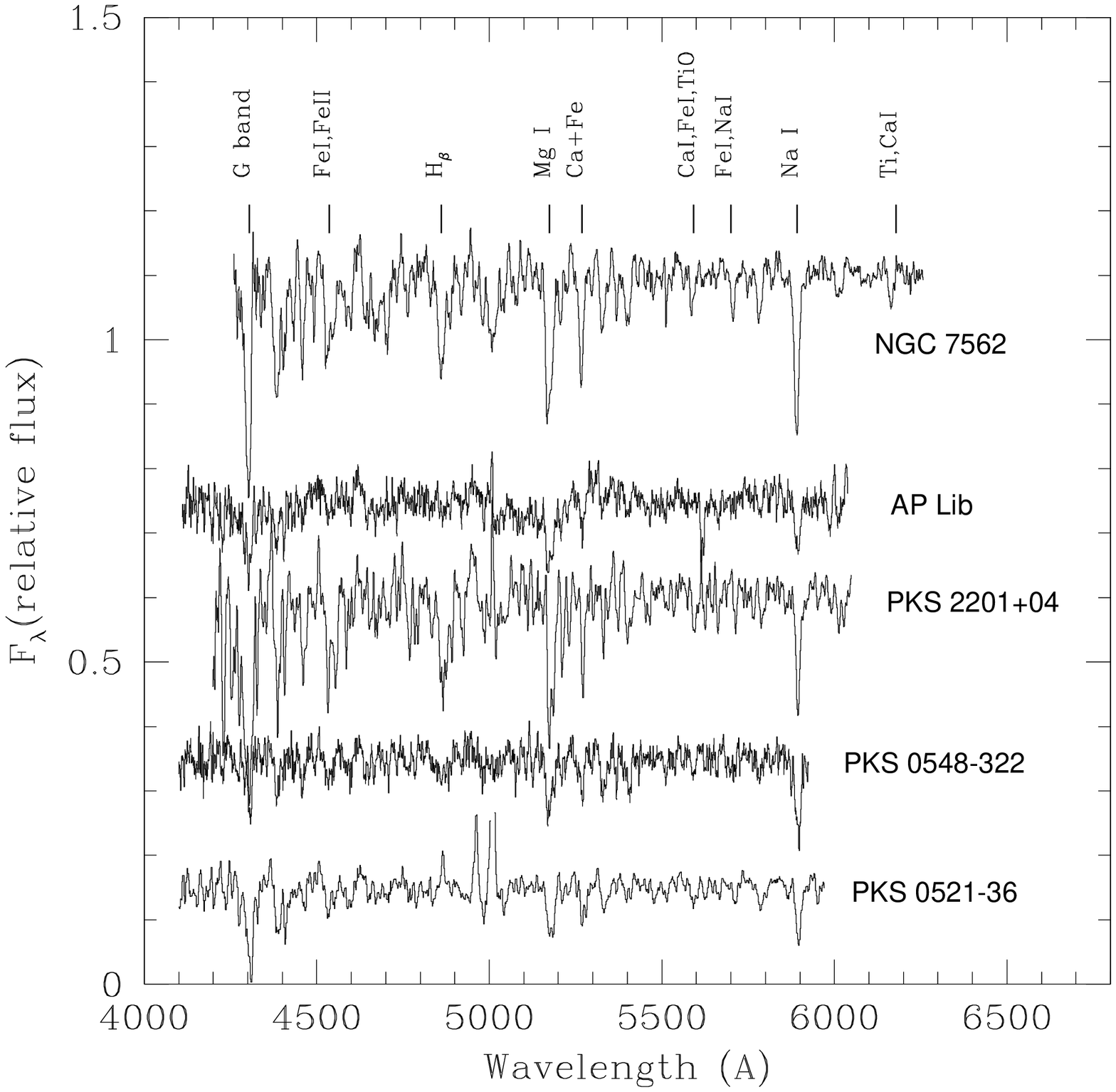}
\caption{The optical rest frame spectra of the BL Lacs observed at ESO compared with that 
of the nearby  elliptical NGC 7562. The spectra have been continuum-normalized 
 to unity. The spectra (from bottom to top) 
have been shifted vertically by -0.85,  -0.65,  -0.4, -0.25 and +0.1 units to 
avoid overlap.
Main absorption features are identified. 
Emission lines are present in the spectrum of all these BL Lac objects but PKS 0548-322.
These emission features have been removed from the analysis of the stellar velocity 
dispersion (see text for details). The strongest emission lines have been  truncated 
to avoid superposition of the graphs. }

\addtocounter{figure}{-1}%
\end{figure}
\begin{figure}
\epsscale{1}
\plotone{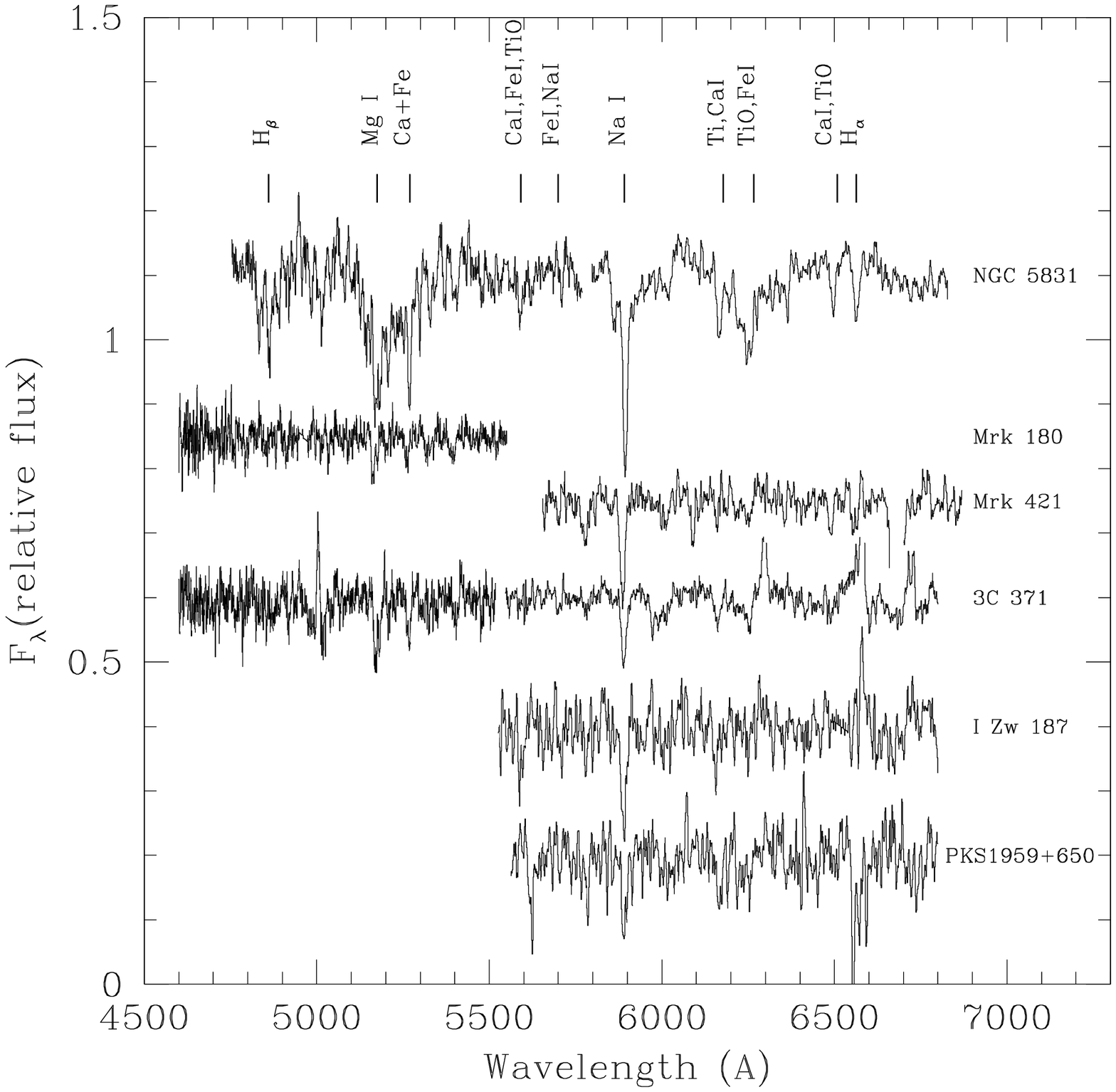}
\caption{Continued. 
The optical spectra of the  BL Lacs (rest frame) observed at the NOT (FKT02) compared with that 
of the nearby  elliptical NGC 5831. The reproduced spectra have been continuum-normalized 
 to unity. The tracings of the spectra (from bottom to top) 
have been shifted vertically by -0.8, -0.6, -0.4, -0.25, -0.15, +0.1. 
 The strongest emission lines have been  truncated 
to avoid superposition of the graphs. 
In the case of Mrk 421 the tracing around the telluric B band was also omitted.
}
\end{figure}

\begin{figure}
\epsscale{1}
\plotone{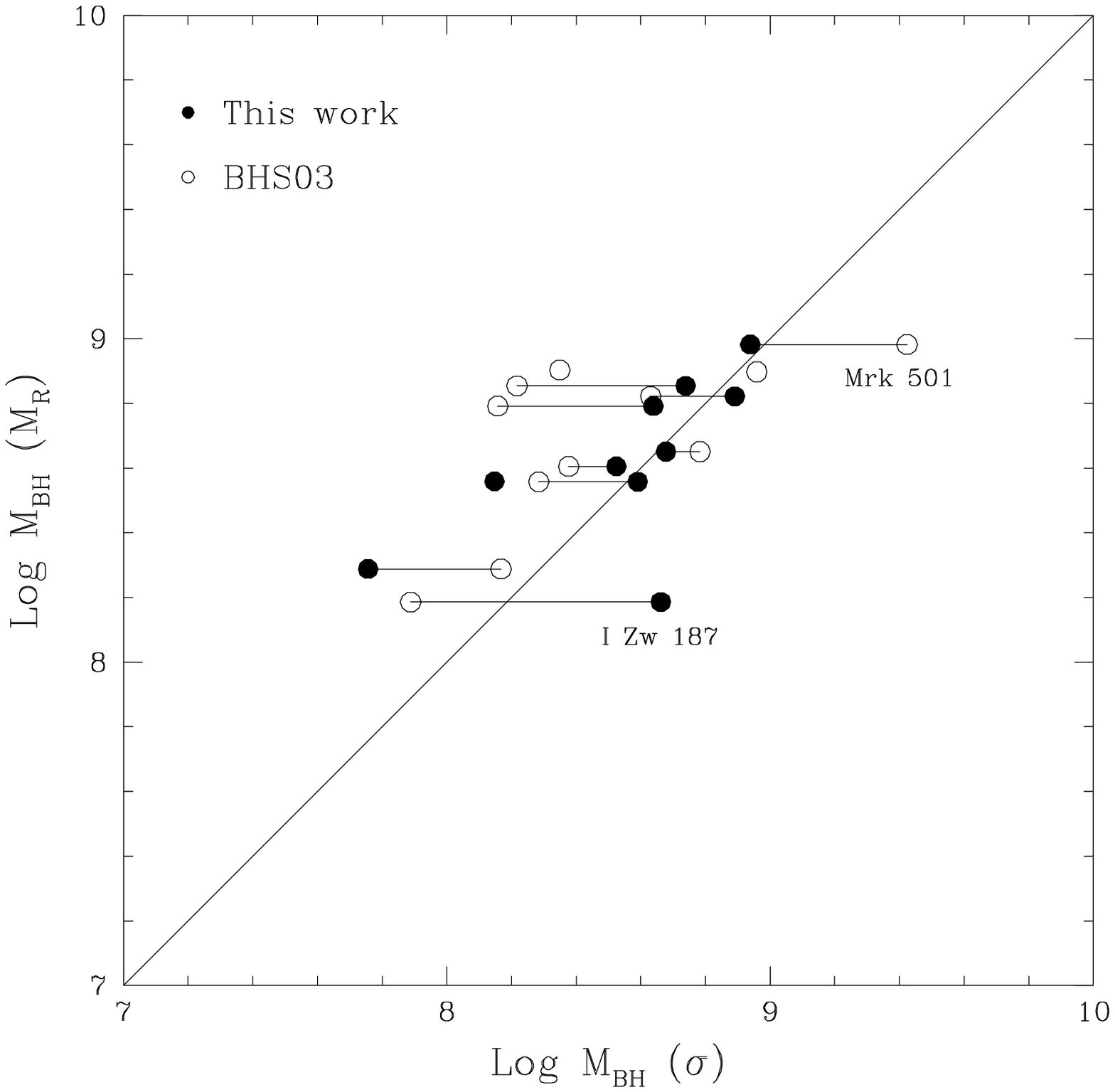}
\caption{Comparison between BH masses derived from the 
M$_{BH}$ -- $\sigma$ and the M$_{BH}$ -- M$_R$(bulge) relations 
(defined by Bettoni et al 2002 ) for the total 
sample of 12 BL Lacs with measured $\sigma$. 
Filled circles refer to this work and FKT02 while 
 open circles refer to BHS03. 
}
\end{figure}

\begin{figure}
\epsscale{1}
\plotone{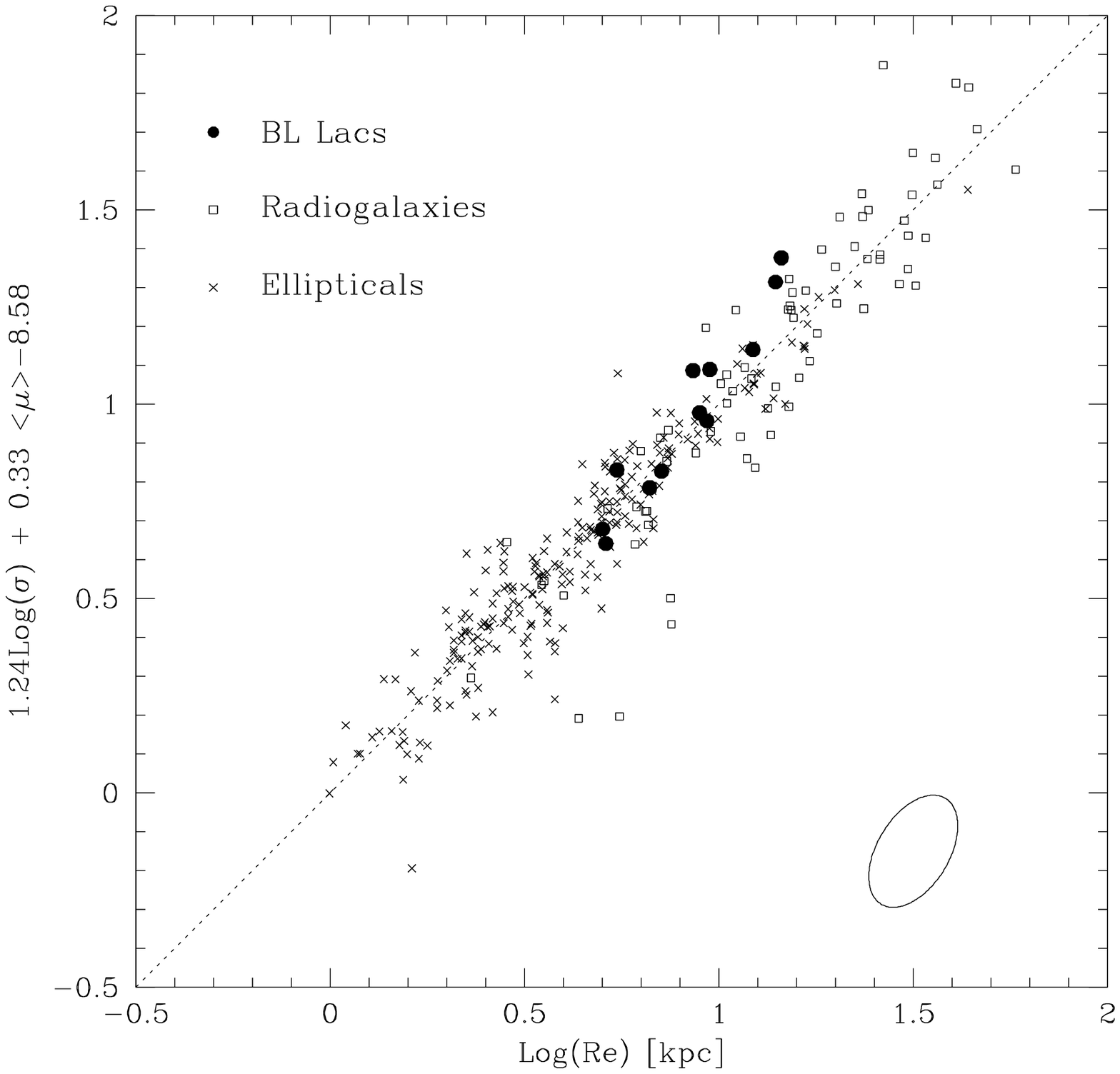}
\caption{The fundamental plane of BL Lacs (filled circles) compared with low redshift 
radio galaxies (open squares; Bettoni et al. 2001) and normal ellipticals 
(crosses; J\"orgensen et al. 1996).
For BL Lacs, the typical error region is indicated by the ellipse.
The size of the minor and major axes are derived from the average error on the quantities plotted 
in the X and Y axes,  respectively. The error for the Y-axis is 
derived combining the average uncertainty on $\sigma$ and $<\mu>$. 
Note that the error of $<\mu>$ in addition to the photometric uncertainty 
($\sim$ 0.1 magnitude) also significantly depends on the error of R$_e$. 
The orientation of the major axis of the ellipse is determined by the 
relation  $<\mu> \propto $ 5$\times$Log(R$_e$). }.
\end{figure}

\clearpage

%

\begin{deluxetable}{llrcrll}
\tablecolumns{7} 
\tablewidth{0pc} 
\tablecaption{Observations and results.}
\tablehead{
\colhead{Object} & \colhead{z} & \colhead{Setup}\tablenotemark{a}  
& \colhead{} 
&\colhead{Exposure time} &\colhead{S/N}\tablenotemark{b}  & \colhead{$\sigma$}\\
\colhead{} & \colhead{} & \colhead{} & \colhead{} & \colhead{(sec)} & \colhead{} & 
\colhead{km s$^{-1}$} }
\startdata
NGC 2986 	& 0.0077 & A  &   & 600 &50  & 260$\pm$10\\
	        &        &  B &  & 600  &55  & 275$\pm$10 \\
NGC 5831 	& 0.0055 & A  &   & 600  &40  & 187$\pm$15 \\
NGC 7562	& 0.012  & A  &   & 600  &45  & 285$\pm$12\\
	        &     &  B &  & 600  &55  & 280$\pm$10 \\
                &        &    &   &      &    &            \\
AP Lib 	& 0.049  &  A &  & 3000 &40  & 250$\pm$12 \\
PKS 0521-36	& 0.055 & A  &   & 3600 &45  & 249$\pm$8 \\
	&             & B  &   & 3000 &45  & 261$\pm$8 \\
PKS 0548-322 & 0.069    & A  &  & 2400 &40  & 269$\pm$10 \\
       	&             & B  &  & 3000 &55  & 257$\pm$10 \\
PKS 2201+04   & 0.027   & A  &  & 3600 & 30 & 165$\pm$15 \\
 \enddata
  \tablenotetext{a}{ Setup A: range 4400 -- 6300 \AA. Setup B: range 6300 -- 8000 \AA}
 \tablenotetext{b}{ S/N per pixel.}
  \end{deluxetable}


\begin{deluxetable}{lllllllrl}

\tablecolumns{9} 
\tablewidth{0pc} 
\tablecaption{Properties of host galaxies and BH masses of BL Lacs
}
\tablehead{
\colhead{Object\tablenotemark{a} } & 
\colhead{T\tablenotemark{b}  } & 
\colhead{$z$} & 
\colhead{$\sigma$\tablenotemark{c} } & 
\colhead{log($M_{BH}$)$_\sigma$\tablenotemark{d}} & 
\colhead{m$_R$\tablenotemark{e}} & 
\colhead{M$_R$\tablenotemark{f}} & 
\colhead{R$_e$\tablenotemark{e, f}} & 
\colhead{log($M_{BH}$)$_{b}$\tablenotemark{g}} \\

\colhead{}  & \colhead{} &  \colhead{} &
\colhead{(km/s)} & 
\colhead{[M$_{\odot}$]} & \colhead{} & \colhead{} & \colhead{kpc} & 
\colhead{[M$_{\odot}$]} 
}
\startdata
Mrk 421   &H & 0.031 & 236 $\pm$ 10&  8.52 $\pm$ 0.13 & 13.17$\pm$ 0.15  & -23.21 &  9.5$\pm$ 3.0 &  8.61 \\ 
Mrk 180   & H & 0.045 & 244  $\pm$ 10& 8.59 $\pm$ 0.13 & 14.09$\pm$ 0.14  & -23.12 &  8.6$\pm$ 2.5 &  8.56 \\ 
Mrk 501   & H & 0.034 & 291  $\pm$ 13& 8.94 $\pm$ 0.15 & 12.62$\pm$ 0.29  & -23.96 & 12.2$\pm$ 3.9 &  8.98 \\ 
I Zw 187$^\dag$ &H* & 0.055 & 171 $\pm$ 12 & 7.89 $\pm$ 0.17 & 15.28$\pm$ 0.07  & -22.37 & 5.5 $\pm$ 0.5&  8.19 \\ 
3C 371     &L*  & 0.051 & 284 $\pm$ 18 & 8.89 $\pm$ 0.17 & 13.84$\pm$ 0.05 & -23.64 & 14.5$\pm$ 1.5 &  8.82 \\ 
1ES 1959+650   &H & 0.048 & 195 $\pm$ 15 &  8.15 $\pm$ 0.17 & 14.23$\pm$ 0.16  & -23.12 &6.6$\pm$ 3.5  & 8.56 \\ 
PKS 2201+04   & L* & 0.027 & 160 $\pm$ 7 &  7.76 $\pm$ 0.13 & 13.50$\pm$ 0.12  & -22.58 &5.1$\pm$ 0.5 &  8.29 \\ 
PKS 0521-36 &L* & 0.055 & 255 $\pm$ 8 &  8.68 $\pm$ 0.12 & 14.35$\pm$ 0.20  & -23.30 & 5.0$\pm$ 2.5 &   8.65 \\ 
PKS 0548-322 & H& 0.069 & 263 $\pm$ 10 &  8.74 $\pm$ 0.13 & 14.45$\pm$ 0.3  & -23.71 &14.0$\pm$ 1.5 &   8.85 \\ 
AP Lib         &H* & 0.049 & 250 $\pm$ 12  & 8.64 $\pm$ 0.14 & 13.81$\pm$ 0.08  & -23.58 & 8.9$\pm$ 1.0  & 8.79 \\ 
0706+591$^\dag$ &H & 0.125 & 216 $\pm$ 23&  8.35 $\pm$ 0.22 & 15.70$\pm$ ...  & -23.81 &9.3$\pm$ ...  & 8.90 \\ 
2344+514$^\dag$ & H& 0.044 & 294 $\pm$ 24 &  8.96 $\pm$ 0.20  & 13.36$\pm$ 0.05  & -23.80 &7.1$\pm$ 1.6  & 8.90 \\ 
 \enddata
\tablenotetext{a}{Objects for which the values of $\sigma$ is taken from Barth et al 2003 
are marked with $^\dag$ } 
\tablenotetext{b}{Type of BL Lac object: L= Low frequency peaked BLL; H = High frequency peaked BLL. \\
An asterisk indicates that the spectrum shows emission features.}
\tablenotetext{c}{Central velocity dispersion corrected to a common aperture as in FKT02}
\tablenotetext{d}{The uncertainty on the BH mass is derived from the combination 
of the errors on $\sigma$ and \\ the dispersion of the M$_{BH}$--$\sigma$ relation.}
\tablenotetext{e}{Total apparent magnitude of the host galaxy corrected for galactic extinction. \\ 
Adopted values are the median of the measurements collected from the literature: see BHS03; 
Falomo et al 1995); \\ Falomo (1996); Falomo \& Kotilainen (1999); Falomo \& Kotilainen (2003); 
Heidt et al 1999 ; \\ 
Nilsson et al (1999); Nilsson et al (2003); Pursimo et al (2002);
Scarpa et al (2000); \\ 
Stickel et al (1993); Wurtz et al (1996) }
\tablenotetext{f}{Parameters are computed assuming H$_0$ = 50 km s$^{-1}$
Mpc$^{-1}$ and $\Omega$ = 0 }
\tablenotetext{g}{BH masses derived from the host galaxy luminosity and equation 2 in the text.}
 \end{deluxetable}

\end{document}